# Imaging Mechanism of Piezoresponse Force Microscopy in Capacitor Structures

Sergei V. Kalinin[*] and Brian J. Rodriguez[†],

Materials Science and Technology Division and the Center for Nanophase Materials Science,

Oak Ridge National Laboratory, Oak Ridge, TN 37831

Seung-Hyun Kim

INOSTEK Inc., Gyeonggi 426-901, Korea

S-K. Hong

Hynix Semiconductor, Inc., San 136-1, Ami-ri, Bubal-eub, Icheon-si,

Kyoungki-do 467-701, Korea

Alexei Gruverman

Department of Physics and Astronomy, University of Nebraska-Lincoln,

Lincoln, NE 68588-0111

Eugene A. Eliseev[‡]

Institute for Problems of Materials Science, National Academy of Science of Ukraine,

3, Krjijanovskogo, 03142 Kiev, Ukraine

---

[*] Corresponding author, sergei2@ornl.gov

[†] Current Address, Max Planck Institute for Microstructural Physics, Weinberg 2, Halle\Saale, Germany 06120

[‡] Corresponding author, eliseev@i.com.ua

The image formation mechanism in Piezoresponse Force Microscopy (PFM) of capacitor structures is analyzed. We demonstrate that the spatial resolution is a bilinear function of film and top electrode thicknesses, and derive the corresponding analytical expressions. For many perovskites, the opposite contributions of $d_{31}$ and $d_{33}$ components can result in anomalous domain wall profiles. This analysis establishes the applicability limits of PFM for polarization dynamics studies in capacitors, and applies to other structural probes, including focused X-ray studies of capacitor structures.

Applications of ferroelectric materials as non-volatile memory components[1,2] has resulted in significant interest in theoretical and experimental studies of polarization dynamics in capacitor structures. Numerous theoretical studies of polarization behavior, depolarization field and dead-layer effects, and periodic domain-structure stabilities using Ginzburg-Landau and atomistic theories are available.[3,4,5] Experimentally, spatially-resolved polarization dynamics have been addressed by focused X-ray measurements[6] and Piezoresponse Force Microscopy (PFM).[7] In PFM, a periodic uniform electric field is applied between the top and bottom electrodes, and the resulting surface displacement is measured locally by a Scanning Probe Microscope (SPM) tip.[8] This global excitation method is different from standard (local excitation) PFM, in which the electric field is applied directly to the tip, and the local strain is detected. The PFM studies of capacitor structures have allowed the stability of uniformly switched states,[9] the flexoelectric contribution to polarization stability,[10] and the localization of nucleation sites[11] and statistical aspects of nucleation to be studied.[12] Despite the significant number of polarization imaging and switching studies in capacitors by PFM, the image formation mechanism, including parameters such as spatial resolution and information limit and the effect of the top electrode on both switching and imaging processes, has not yet been addressed. Here, we analyze experimentally and theoretically the spatial resolution of PFM with global excitation, and predict the implications for PFM-based capacitor studies.

The measurements are performed on a specially designed capacitor structure shown in Fig. 1 (a). The central capacitor pad of ~150 nm thickness is fabricated on the patterned bare film, allowing direct comparison of the PFM resolution within the electroded area and the free surface. Additionally, measurements were performed using a series of different capacitors with varying top electrode thickness. PFM was implemented on a commercial Veeco

MultiMode NS-IIIA SPM system using Au coated tips (NSC-36 B, Micromasch, resonant frequency ~ 155 kHz, spring constant $k$ ~ 1.75 N/m). Thickness studies were performed using a Park Scientific Instruments Autoprobe M5.

Data in Fig. 1 (b,c) illustrates the presence of as-grown domains on the order of the grain size with the characteristic domain wall width of 10 nm on the free surface. In comparison, in the electroded region, much larger than grain-size domains, with characteristic wall widths on the order of 100 nm are formed in as-fabricated capacitor structures. Switched capacitor structures [Fig. 1 (e,f)] exhibit remanent domains that are also larger than the grain-size. Similar behavior is observed in other capacitor structures.

Here, we analyze the domain wall width in PFM on capacitor structures, i.e. the resolution limit in PFM with global excitation. We utilize the decoupled approximation[13,14] that has previously been used for analysis of PFM with local excitation.[15,16] In the geometry of the capacitor experiment, the vertical PFM signal, i.e., the surface displacement $u_{3i}(\mathbf{x})$ at location $(x_1, x_2, 0)$ is :

$$u_3(x_1, x_2, 0) = \int_{-\infty}^{\infty} d\xi_1 \int_{-\infty}^{\infty} d\xi_2 \int_{L}^{L+H} d\xi_3 \frac{\partial G_{3j}(-\xi_1, -\xi_2, \xi_3)}{\partial \xi_k} c_{kjmn} E_3 d_{3nm}(x_1 + \xi_1, x_2 + \xi_2, \xi_3) \quad (1)$$

The coordinates $\mathbf{x} = (x_1, x_2, x_3)$ are linked to the sample [Fig. 2 (a)]. Coefficients $d_{mnk}$ are position dependent components of the piezoelectric strain tensor, $c_{jlmn}$ are elastic stiffness constants tensors and $E_k$ is the external electric field components. The Green's function for a semi-infinite medium $G_{3j}(\mathbf{x}, \xi)$ links the eigenstrains $c_{jlmn} d_{mnk} E_k$ to the displacement field. Here, we use an elastically isotropic approximation for mechanical properties.[17,18]:

Considering the domain structure with two opposite domains divided by flat wall perpendicular to the film surface, we obtain:

$$d_{inm}(\xi_1,\xi_2,\xi_3) = \begin{cases} 0, & \xi_3 > L+H \cup \xi_3 < L; \\ -d_{i\alpha}, & L < \xi_3 < L+H \cap \xi_1 < 0; \\ d_{i\alpha}, & L < \xi_3 < L+H \cap \xi_1 > 0; \end{cases} \quad (2)$$

Using the parity conditions of integrands and Eq. (2), Eq. (1) can be rewritten as follows

$$u_3(x_1,x_2,0) = -U\big(W_{33}(x_1,L,H)d_{33} + (1+2\nu)W_{31}(x_1,L,H,\nu)d_{31}\big) \quad (3)$$

Here $U = HE_3$ is the potential difference between top and bottom electrode. After cumbersome integrations, functions $W_{33}$ and $W_{31}$ are derived analytically as (see Appendix A):

$$W_{33}(x_1,L,H) = f_1(x_1,L,H) + f_2(x_1,L,H) \quad (4a)$$

$$W_{31}(x_1,L,H,\nu) = f_1(x_1,L,H) + \frac{\nu}{2(1+2\nu)} f_2(x_1,L,H) \quad (4b)$$

where

$$f_1(x_1,L,H) = \frac{2}{\pi}\left(\left(\frac{L}{H}+1\right)\arctan\left(\frac{x_1}{L+H}\right) - \left(\frac{L}{H}\right)\arctan\left(\frac{x_1}{L}\right)\right) \quad (4c)$$

$$f_2(x_1,L,H) = \frac{2}{\pi}\frac{x_1}{H}\ln\left(\frac{(L+H)^2 + x_1^2}{L^2 + x_1^2}\right). \quad (4d)$$

Far from the wall, $x_1 \to \pm\infty$, functions Eq. (4a,b) tend to $W_{3i}(x_1,L,H) \to \pm 1$. Hence, the piezoresponse signal in a uniformly poled capacitor is $u_3(x_1,x_2,0)/U = -d_{33} - (1+2\nu)d_{31}$. The non-zero contribution of $d_{31}$ is directly related to the mechanical conditions on the boundary between film and bottom electrode, which are assumed to have same mechanical properties. The dependence of functions $W_{3i}$ on coordinate $x_1$ is shown on Fig. 3 (a). Note that the component profile width is much higher for $W_{31}$ than for $W_{33}$, which is related to the smaller contribution of $f_2$ to $W_{31}$. As the consequence of the different widths of $W_{31}$ and $W_{33}$, full PFM response profile (3) exhibits an anomalous shape with minimum and maximum near the wall center for the most encountered case of $d_{31}/d_{33} \sim -0.5 \div -0.3$ (see Fig. 3(c,d)). At the same time, the slope of the response at the wall center is determined primarily by $d_{33}$. With $d_{31}$ absolute value increase, the saturation level decreases much faster than the slope, so that the width also decreases.

Here, we define the generalized domain wall width, or PFM resolution, as the distance between the points at which the response is saturated to the fraction, $\eta$, of signal at infinity. The width can be approximated as a bilinear function, $w(L,H) \cong aH + bL$, of the film and top electrode thicknesses,

$$w(L,H) \cong a\left(\eta, v, \frac{d_{31}}{d_{33}}\right) H + b\left(\eta, v, \frac{d_{31}}{d_{33}}\right) L \tag{5a}$$

where coefficient, $a$, is the solution of the transcendental equation:

$$\arctan\left(\frac{a}{2}\right) + \frac{d_{33} + v d_{31}/2}{d_{33} + (1+2v)d_{31}} \ln\left(1 + \frac{4}{a^2}\right)\frac{a}{2} = \frac{\pi}{2}\eta, \tag{5b}$$

and coefficient $b$ is

$$b = \eta \frac{\pi}{2} \frac{d_{33} + (1+2v)d_{31}}{d_{33} + v d_{31}/2}. \tag{5c}$$

Typical values for coefficient $a$ varies from 0.08 to 0.4 at $\eta=0.76$ and $-d_{33}/d_{31}=2\div 4$ (see Table I). The dependence of width on top electrode thickness obtained by the numerical calculation and with the help of Eqs. (5) is shown in Fig. 2 (b). The immediate consequence of Eq. (5a) is that for a mechanically uniform capacitor (no cracks, etc), the signal generation area in PFM even for point mechanical contact is determined by the total thickness of the capacitor structure. Note that domain wall width scales linearly with sample thickness for thin top electrodes ($W/H$ = 0.08 – 0.3 depending on piezoelectric anisotropy), and increases rapidly with top electrode thickness. This agrees with experimental data obtained at several capacitor thicknesses summarized in Table II for (111) oriented 180 nm thick PZT film, and reported elsewhere.[19] Note that this conclusion can readily be understood as a consequence of the presence of the primary length scale in the system, and similar behavior is anticipated for e.g., synchrotron X-ray based measurements of domain structures.

For cases when the domain wall is also associated with the grain boundary and the grains are mechanically decoupled, the coefficient in Eq. (5) is $a = 0$. Here, the top electrode thickness is the only relevant length scale. For the case of intermediate coupling, the bulk contribution will be reduced, but the top-electrode contribution is expected to be constant. Finally, for liquid top electrode (imaging in conductive solution), $b = 0$.

To summarize, we have analyzed the imaging mechanism in PFM with global excitation. The analytical expressions for the domain wall profile are derived. The resolution, i.e. the measured domain wall profile is derived for an infinitely thin wall and a point mechanical contact approximation. It is shown that domain wall width scales linearly with the thickness of the ferroelectric layer and top electrode, albeit with significantly different slopes. For typical material parameters and thin top electrode ($L<<H$), the resolution is expected to be $w \sim 0.2\ H$, which presents the ultimate limit on PFM resolution in capacitors imposed by the presence of the characteristic length scale. Similar behavior is anticipated for other probes of domain wall, including focused X-ray. Higher resolutions can be observed only if walls are mechanically uncoupled, or as a result of the cross-talk between topography and PFM signal, unrelated to domain structures.

National Laboratory, managed and operated by UT-Battelle, LLC. One of the authors (BJR) acknowledges the support of the Alexander von Humboldt Foundation.8

**Appendix A.**

In the decoupled problem the electric field in the material is calculated using a rigid electrostatic model (no piezoelectric coupling); the strain or stress field is calculated using constitutive relations for a piezoelectric solid, and the displacement field is evaluated using an appropriate Green's function for an isotropic or anisotropic solid. In this approximation, PFM signal, i.e., surface displacement $u_i(\mathbf{x})$ at location $(x_1, x_2, 0)$ is given by

$$u_i(x_1,x_2,0) = \int_{-\infty}^{\infty} d\xi_1 \int_{-\infty}^{\infty} d\xi_2 \int_0^{\infty} d\xi_3 \frac{\partial G_{ij}(x_1-\xi_1, x_2-\xi_2, \xi_3)}{\partial \xi_k} c_{kjmn} E_l d_{lnm}(\xi_1,\xi_2,\xi_3) \qquad (A.1)$$

Here coordinates $\mathbf{x} = (x_1, x_2, x_3)$ are linked to the sample (Fig. 2). Coefficients $d_{mnk}$ are position dependent components of the piezoelectric strain tensor, $c_{jlmn}$ are elastic stiffness constants tensors and $E_k$ is the external electric field components. The Green's function for a semi-infinite medium $G_{3j}(\mathbf{x},\xi)$ links the eigenstrains $c_{jlmn} d_{mnk} E_k$ to the displacement field.

For most inorganic ferroelectrics, the elastic properties of material are weakly dependent on orientation. Hence, material can be approximated as elastically isotropic. Corresponding Green's tensor for elastic isotropic half-plane is given by Lurie[17] and Landau and Lifshitz:[18]

$$G_{ij}(x_1-\xi_1, x_2-\xi_2, \xi_3) = \begin{cases} \frac{1+\nu}{2\pi Y}\left[\frac{\delta_{ij}}{R} + \frac{(x_i-\xi_i)(x_j-\xi_j)}{R^3} + \frac{1-2\nu}{R+\xi_3}\left(\delta_{ij} - \frac{(x_i-\xi_i)(x_j-\xi_j)}{R(R+\xi_3)}\right)\right] & i,j \neq 3 \\[6pt] \frac{(1+\nu)(x_i-\xi_i)}{2\pi Y}\left(\frac{-\xi_3}{R^3} - \frac{(1-2\nu)}{R(R+\xi_3)}\right) & i=1,2 \text{ and } j=3 \\[6pt] \frac{(1+\nu)(x_j-\xi_j)}{2\pi Y}\left(\frac{-\xi_3}{R^3} + \frac{(1-2\nu)}{R(R+\xi_3)}\right) & j=1,2 \text{ and } i=3 \\[6pt] \frac{1+\nu}{2\pi Y}\left(\frac{2(1-\nu)}{R} + \frac{\xi_3^2}{R^3}\right) & i=j=3 \end{cases}$$

(A.2)

where $R = \sqrt{(x_1 - \xi_1)^2 + (x_2 - \xi_2)^2 + \xi_3^2}$, $Y$ is Young's modulus, and $\nu$ is the Poisson ratio. Stiffness tensor $c_{kjmn}$ of the elastically isotropic medium has the view

$$c_{klmn} = \frac{Y}{2(1+\nu)}\left[\frac{2\nu}{1-2\nu}\delta_{kl}\delta_{mn} + \delta_{km}\delta_{ln} + \delta_{kn}\delta_{lm}\right]. \quad (A.3)$$

Here we calculate surface displacement vertical component:

$$u_3(x_1, x_2, 0) = \int_{-\infty}^{\infty} d\xi_1 \int_{-\infty}^{\infty} d\xi_2 \int_{L}^{L+H} d\xi_3 \frac{\partial G_{3j}(-\xi_1, -\xi_2, \xi_3)}{\partial \xi_k} c_{kjmn} E_3 d_{3nm}(x_1 + \xi_1, x_2 + \xi_2, \xi_3) \quad (A.4)$$

The contribution of components $d_{333} \equiv d_{33}$ is determined by the function

$$\widetilde{W}_{33}(\xi_1, \xi_2, \xi_3) = \frac{\partial G_{3j}(-\xi_1, -\xi_2, \xi_3)}{\partial \xi_k} c_{kj33} = -\frac{3}{2\pi}\frac{\xi_3^3}{R^5}, \quad (A.5a)$$

while $d_{311} = d_{322} \equiv d_{31}$ contribution is

$$\widetilde{W}_{31}(\xi_1, \xi_2, \xi_3) = \frac{\partial G_{3j}(-\xi_1, -\xi_2, \xi_3)}{\partial \xi_k} c_{kj11} + \frac{\partial G_{3j}(-\xi_1, -\xi_2, \xi_3)}{\partial \xi_k} c_{kj22} = \frac{1}{2\pi}\frac{\xi_3}{R^3}\left(3\frac{\xi_3^2}{R^2} - 2(1+\nu)\right).$$

(A.5b)

Here $R = \sqrt{\xi_1^2 + \xi_2^2 + \xi_3^2}$. Considering the simple domain structure with two opposite domains divided by flat wall, perpendicular to the film surface, we suppose the following dependence on the coordinates:

$$d_{inm}(\xi_1, \xi_2, \xi_3) = \begin{cases} 0, & \xi_3 > L+H \cup \xi_3 < L; \\ -d_{i\alpha}, & L < \xi_3 < L+H \cap \xi_1 < 0; \\ d_{i\alpha}, & L < \xi_3 < L+H \cap \xi_1 > 0; \end{cases} \quad (A.6)$$

Using the parity conditions of integrands and evident dependence (A.6), one can rewrite (A.4) as follows

$$u_3(x_1,x_2,0) = 4\int_0^{x_1} d\xi_1 \int_0^{\infty} d\xi_2 \int_L^{L+H} d\xi_3 \left(\tilde{W}_{33}(\xi_1,\xi_2,\xi_3)E_3 d_{33} + \tilde{W}_{31}(\xi_1,\xi_2,\xi_3)E_3 d_{31}\right) =$$
$$= -W_{33}(x_1,L,H)HE_3 d_{33} - (1+2\nu)W_{31}(x_1,L,H)HE_3 d_{31}$$

(A.7)

Here the combination $HE_3$ is the potential difference between top and bottom electrode.

Using the definition (A.7) one can integrate in the cylindrical system of coordinates on radial and axial coordinates, which leads to one-fold integral on polar angle

$$W_{33}(x_1,L,H) = -\frac{1}{H} 4\int_0^{x_1} d\xi_1 \int_0^{\infty} d\xi_2 \int_L^{L+H} d\xi_3 \tilde{W}_{33}(\xi_1,\xi_2,\xi_3) =$$

$$= \text{sign}(x_1)\left(1 - \frac{2}{\pi H}\int_0^{\pi/2} \frac{d\varphi}{\cos\varphi}\left(\frac{(L+H)^2\cos^2\varphi + 2x_1^2}{\sqrt{(L+H)^2\cos^2\varphi + x_1^2}} - \frac{L^2\cos^2\varphi + 2x_1^2}{\sqrt{L^2\cos^2\varphi + x_1^2}}\right)\right) =$$

(A.8a)

and after lengthy manipulations one can get the following:

$$W_{33}(x_1,L,H) =$$
$$= \left(\text{sign}(x_1) - \frac{2}{\pi}\left(\frac{x_1}{H}\ln\left(\frac{L^2 + x_1^2}{(L+H)^2 + x_1^2}\right) + \left(\frac{L}{H}+1\right)\arctan\left(\frac{L+H}{x_1}\right) - \frac{L}{H}\arctan\left(\frac{L}{x_1}\right)\right)\right)$$

(A.8b)

Calculations for $d_{31}$ contribution:

$$W_{31}(x_1,L,H) = -\frac{1}{H(1+2\nu)} 4\int_0^{x_1} d\xi_1 \int_0^{\infty} d\xi_2 \int_L^{L+H} d\xi_3 \tilde{W}_{31}(\xi_1,\xi_2,\xi_3) = -\frac{W_{33}(x_1,L,H)}{(1+2\nu)} +$$

$$+ 2\frac{(1+\nu)}{(1+2\nu)}\text{sign}(x_1)\left(1 - \frac{2}{\pi H}\int_0^{\pi/2} \frac{d\varphi}{\cos\varphi}\left(\sqrt{(L+H)^2\cos^2\varphi + x_1^2} - \sqrt{L^2\cos^2\varphi + x_1^2}\right)\right) =$$

$$= -\frac{2}{\pi(1+2\nu)}\left(-\frac{x_1}{H}\ln\left(\frac{L^2+x_1^2}{(L+H)^2+x_1^2}\right) + \left(\frac{L}{H}+1\right)\arctan\left(\frac{x_1}{L+H}\right) - \left(\frac{L}{H}\right)\arctan\left(\frac{x_1}{L}\right)\right) +$$

$$+ \frac{2(1+\nu)}{(1+2\nu)}\frac{2}{\pi}\left(-\frac{x_1}{2H}\ln\left(\frac{L^2+x_1^2}{(L+H)^2+x_1^2}\right) + \left(\frac{L}{H}+1\right)\arctan\left(\frac{x_1}{L+H}\right) - \left(\frac{L}{H}\right)\arctan\left(\frac{x_1}{L}\right)\right) =$$

$$= \frac{2}{\pi}\left(-\frac{\nu}{(1+2\nu)}\frac{x_1}{2H}\ln\left(\frac{L^2+x_1^2}{(L+H)^2+x_1^2}\right) + \left(\left(\frac{L}{H}+1\right)\arctan\left(\frac{x_1}{L+H}\right) - \frac{L}{H}\arctan\left(\frac{x_1}{L}\right)\right)\right)$$

(A.9)

Eqs. (A.8b) and (A.10) for the functions $W_{33}$ and $W_{31}$ can be written as

$$W_{33}(x_1, L, H) = \frac{2}{\pi}\left(\left(\frac{L}{H}+1\right)\arctan\left(\frac{x_1}{L+H}\right) - \left(\frac{L}{H}\right)\arctan\left(\frac{x_1}{L}\right) - \frac{x_1}{H}\ln\left(\frac{L^2 + x_1^2}{(L+H)^2 + x_1^2}\right)\right)$$

(A.10a)

$$W_{31}(x_1, L, H) =$$
$$= \frac{2}{\pi}\left(\left(\left(\frac{L}{H}+1\right)\arctan\left(\frac{x_1}{L+H}\right) - \frac{L}{H}\arctan\left(\frac{x_1}{L}\right)\right) - \frac{\nu}{(1+2\nu)}\frac{x_1}{2H}\ln\left(\frac{L^2 + x_1^2}{(L+H)^2 + x_1^2}\right)\right)$$

(A.10b)

**Table I.**

Width derivatives on $H$ and $L$, $w = aH + bL$, for $\eta=0.76$, $\nu=0.3$.

| Material | $d_{33}$ (pm/V) | $d_{31}$ (pm/V) | $a$ | $b$ |
|---|---|---|---|---|
| BaTiO$_3$ | 86 | -35 | 0.17 | 0.45 |
| PbTiO$_3$ | 117 | -25 | 0.43 | 0.81 |
| PZT6B | 75 | -29 | 0.19 | 0.49 |
| (111) PZT40/60[20] | 73.5 | -24 | 0.26 | 0.60 |
| LiNbO$_3$ | 6 | -1 | 0.51 | 0.90 |
| LiTaO$_3$ | 8 | -2 | 0.37 | 0.74 |

**Table II**

Experimental and theoretical domain wall width for different top electrode thicknesses

| TE Thickness | Poled cap | Unpoled cap | Theory |
|---|---|---|---|
| 50 | 130-170 | 40-60 | 81 |
| 150 |  | 80 | 150 |
| 250 | 180-250 |  | 217 |

**Figure captions**

**Fig. 1.** (a) Surface topography, and (b) amplitude and (c) phase PFM images of poled thin film. (c) Surface topography, and (d) amplitude and (e) phase PFM images of etched capacitor structure.

**Fig. 2.** (Color online) Coordinate systems in global excitation PFM experiment. (b) Domain wall width on level $\eta=0.76$ as function of the top electrode thickness for $\nu=0.3$ and $-d_{33}/d_{31}=2, 3, 4$ (curves 1, 2, 3). Solid curves represent the numerical calculations, dashed ones are after approximation (5).

**Fig. 3.** (Color online) Coordinate dependence of (a) $d_{31}$ and (b) $d_{33}$ contribution to the vertical PFM response for $\nu=0.3$ and different thickness of top electrode L/H=0, 1, 2, 3 (solid, dotted, dashed and dash-dotted curves). Profiles of the vertical PFM response across the domain wall for $\nu=0.3$, $-d_{33}/d_{31}=2.5, 5$ ((c) and (d)) and different thickness of top electrode L/H=0, 1, 2, 3 (solid, dotted, dashed and dash-dotted curves).

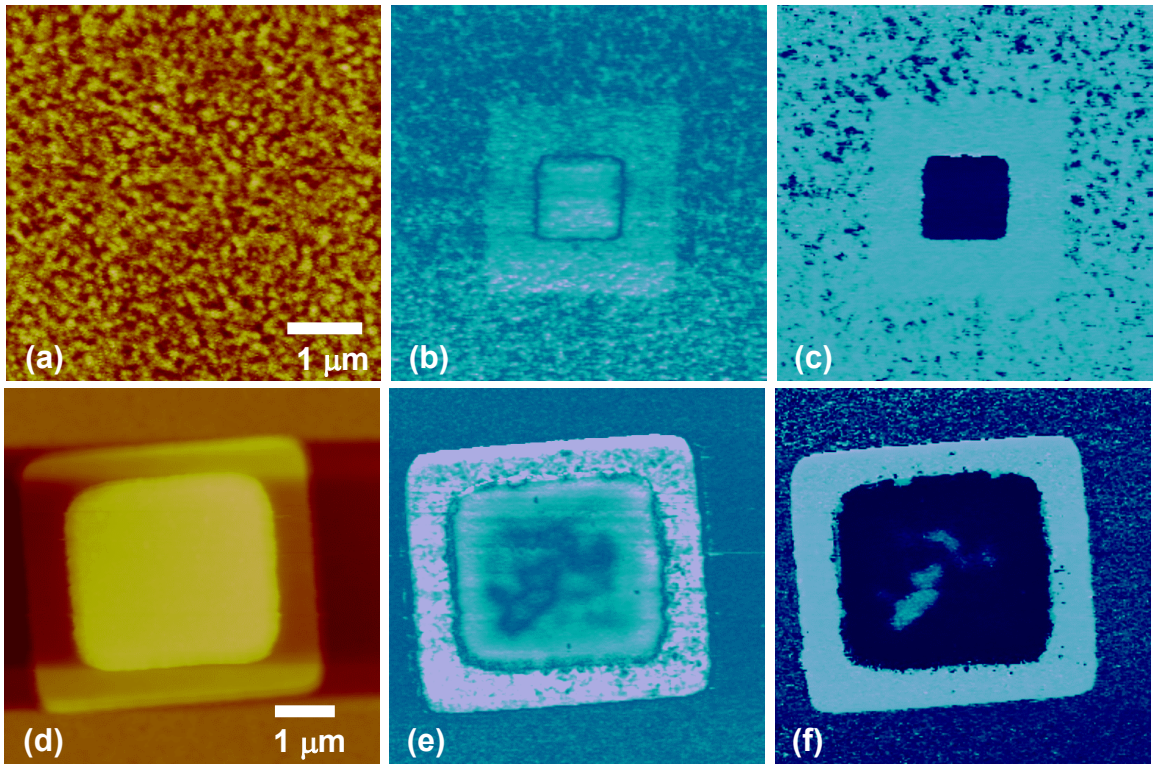

Figure 1. S.V. Kalinin et al

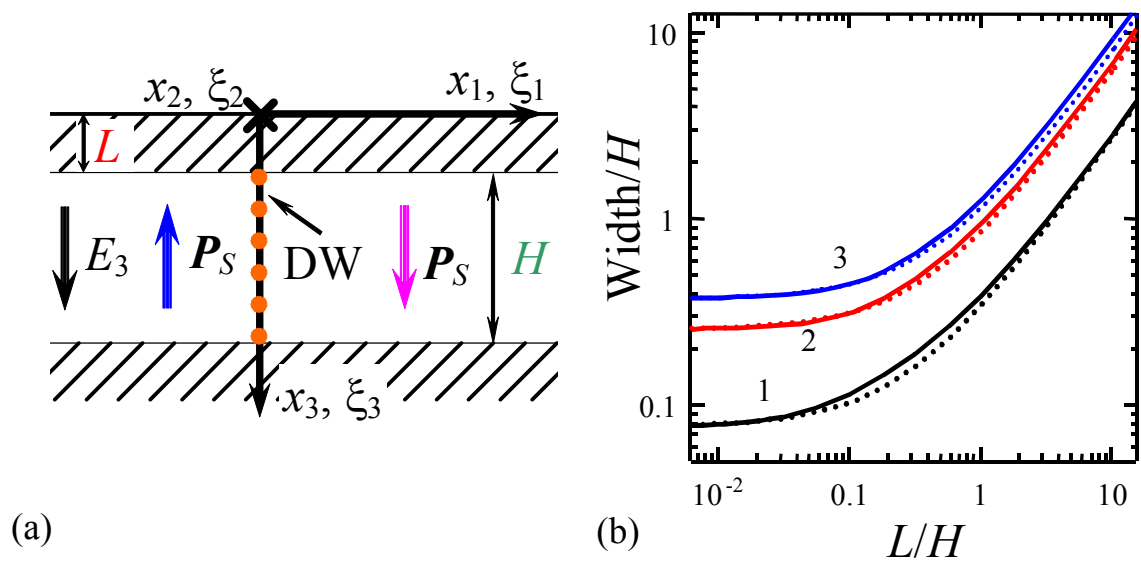

(a)  (b)

Figure 2. S.V. Kalinin et al

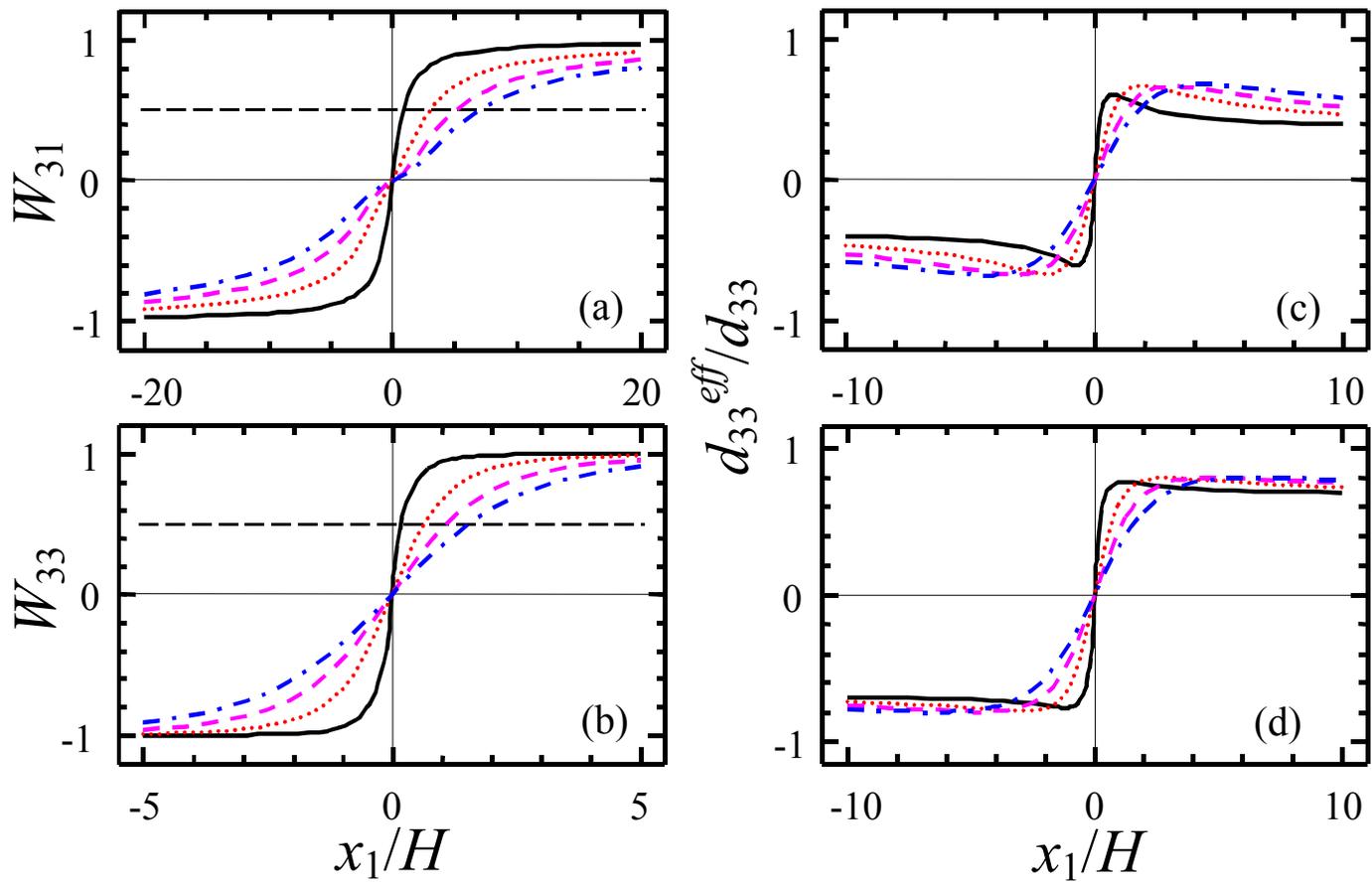

Figure 3. S.V. Kalinin et al

[16] A.N. Morozovska, E.A. Eliseev, S.L. Bravina, and S.V. Kalinin Phys. Rev. **B 75**, 174109 (2007).

[17] A.I. Lur'e, *Three-dimensional problems of the theory of elasticity*. (Interscience Publishers, 1964).

[18] L.D. Landau and E.M. Lifshitz, *Theory of Elasticity*. Theoretical Physics, Vol. 7 (Butterworth-Heinemann, Oxford, 1976).

[19] T. Jungk, A. Hoffmann, and E. Soergel, J. Appl. Phys. **102**, 084102 (2007).

[20] Piezoelectric modules were recalculated from values for (100) oriented PZT40/60 given by M.J. Haun, E. Furman, S.J. Jang, and L.E. Cross, Ferroelectrics **99**, 63(1989).
20